# A Sky Brightness Model for the Starlink 'Visorsat' Spacecraft - II

Richard E Cole[1]

Version 3 – 10th July 2021

## 1 Abstract

A model of the brightness of the 'visorsat' Starlink spacecraft is presented based on published information on the engineering design and from analysis of 131 observations of individual visorsats in late 2020. Comments are offered on the implications of this model on the visibility of visorsat spacecraft across the sky. This is an updated and expanded version of analysis published in Research Notes of the AAS (Cole 2020a). An additional section has been added in this version to consider observations made in June 2021 which indicate brighter visorsat magnitudes.

## 2 Background

Since mid-2019 the launches of SpaceX's Starlink spacecraft have caused considerable and real concern in the astronomical community (Hainaut and Williams 2020). As part of a programme to reduce the brightness of Starlinks, SpaceX launched a single 'Darksat' that used a surface treatment on brighter parts of the spacecraft and more recently developed the 'Visorsat' concept. SpaceX have launched around 1000 visorsats to date (with plans for many more), each with a lightweight, deployable sun-shield or 'visor'. This visor prevents sunlight falling directly onto the white antenna panels on the base of the spacecraft and hence reduces the brightness of the spacecraft as seen from the ground (Figure 1)

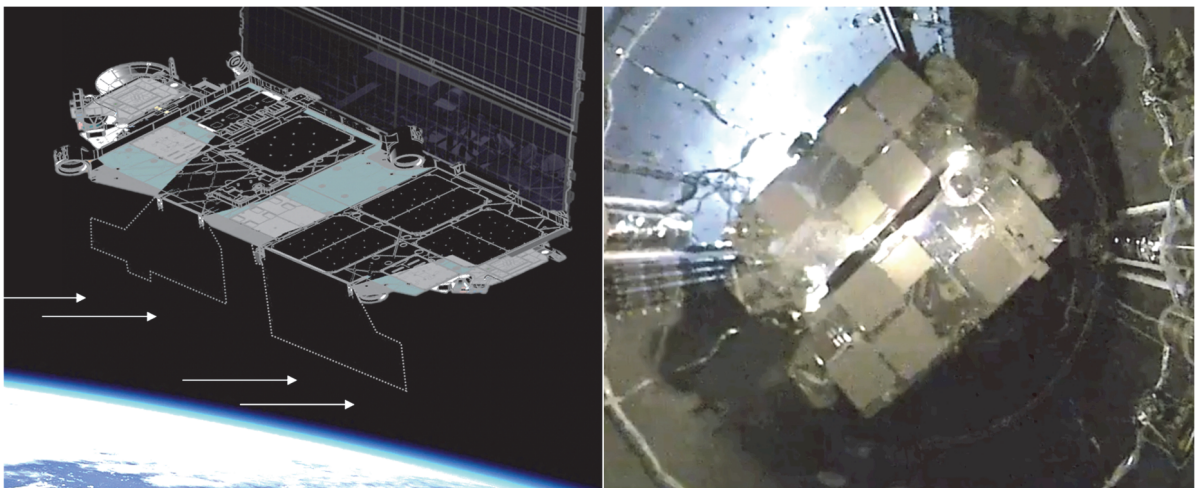

*Figure 1: The 'Visorsat' design (SpaceX 2020) and images of non-visorsat starlinks during the L1.7 launch showing the pale/white antenna panels. Note the base of the spacecraft (facing the camera) is reflecting the inner surface of the Falcon 9 fairing indicating it is manufactured of polished metal.*

[1] St Tudy, Cornwall, United Kingdom

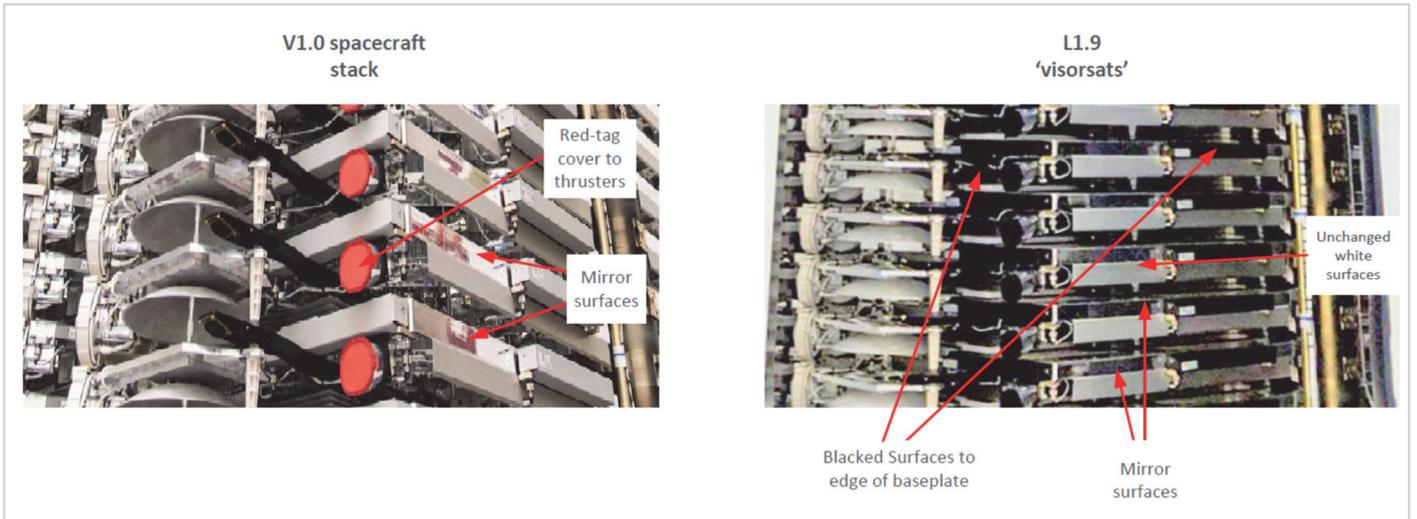

*Figure 2: A comparison of a stack of pre-visorsats on the left and L1.9 visorsats on the right that demonstrate addition of black surfaces to the edges of the bases of each spacecraft (images SpaceX)*

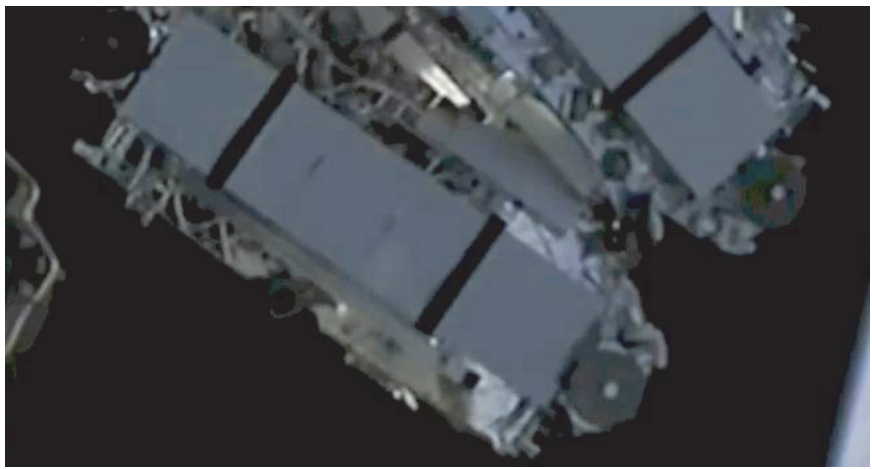

*Figure 3: A image of L1.11 Starlinks just after separation in-orbit. Note the blacked parabolic antennas (in a stowed position) (Images SpaceX)*

Further darkening treatments seem to have been applied to the edges of the spacecraft and to the parabolic antennas (Figure 2 and Figure 3).

The Starlink design includes a large solar-panel. It is this panel that makes the 'trains' of Starlink spacecraft at low altitudes so bright and triggered a widespread concern (and even fear) of how the Starlink constellation could affect the night-sky. More recently, SpaceX have changed the orientation of some of the spacecraft at low altitudes so these panels are not seen as so bright, but the sheer number of spacecraft in a train still has an impact.

When the spacecraft reach their operational altitude of 550km their orientation is changed so that the base and its antennas face the Earth and the solar-panel faces towards the Sun. This configuration is termed 'shark-fin' mode by SpaceX (Figure 4). The visibility from the ground of the solar-panel (blue in this image) in shark-fin mode has not been much considered to date in published works but it is mentioned in SpaceX's account of the visibility reduction methods (SpaceX 2020).

This report does not consider the brightness of Starlinks in non-operational orbits below 550km. Some comments are offered on the appearance of the earlier, non-visorsat spacecraft in the operational orbit.

## 3 The Brightness Model

### 3.1 Observations used

A mathematical model of the brightness of the visorsat design in the operational orbit has been developed and its predictions compared with available observations of visorsats.

Observations – The majority of the observations used here have been acquired by visual observers (Mallama 2020a & 2020b, Respler 2020) and by the author. Other data have been acquired from the on-line database of the MMT telescope (Beskin et al. 2017) and the report of the Dark and Quiet Skies for Science and Society (Dark and Quiet Skies 2020). Given the variety of data sources and observational methods used, the errors are not uniform but in general a magnitude measurement error of +/-0.5$^m$ was deemed appropriate.

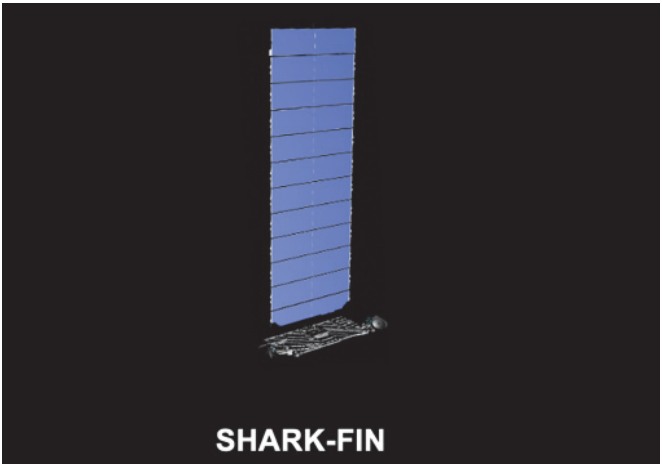

*Figure 4: SpaceX diagram of Shark-fin mode (SpaceX 2020)*

Mathematical models of the Starlink spacecraft in the operational orbit have been developed previously (Mallama 2020c, Dark and Quiet Skies 2020) but these did not include specific consideration of light reflected by the solar-panel. The model described here adds the solar-panel and considers the possible angles at which it is deployed. Also, the detailed effect of the visor on the brightness of the base of the visorsat is considered.

## 3.2 Elements of the model

The solar-panel is modelled as a folded flat surface to replicate the way the separate sections are deployed on-orbit. Images of Starlinks taken from the ground (Vandebergh 2020a) indicate the panel sections deviate significantly from a single plane surface. There may also be small gaps between the panel sections. SpaceX has indicated that the panel is capable of being controlled in its pointing direction via a single motorised axis at the base of the panel. Further, it might be that the panel is offset from the Sun by some angle or offset from the nadir-zenith direction by a different angle (Figure 5).

SpaceX are very aware of the potential visibility of the panel from the ground and the need to control it when the Starlinks are visible from the ground (usually a limited part of the orbit). It is assumed here that the panel control mode fixes the offset angle of the solar-panel (in these parts of the orbit) to minimise visibility; only one offset angle of the panel is considered in analysis of the observations.

Further, the base of the spacecraft can partially or fully block the view of the solar-panel from an observer on the ground. The model includes an approximation of that blocking effect (Figure 5) which is a function of the view angles of the spacecraft by the observer. This blocking is vital to control the brightness of the spacecraft, as discussed later.

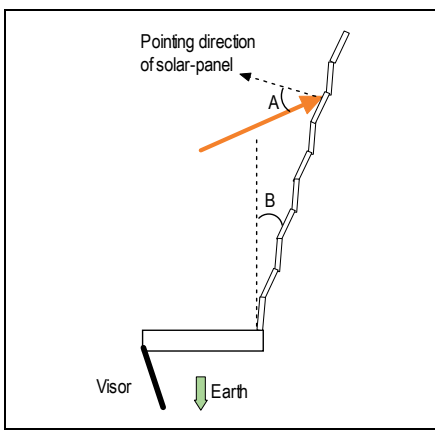
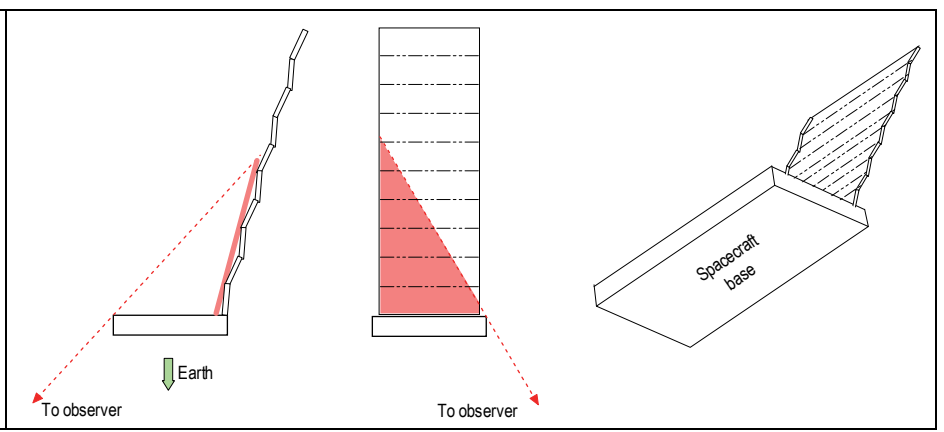

*Figure 5: The Starlink configuration considered in this model. The solar-panel may be oriented at some angle to the Sun (angle A) or fixed at some angle to the nadir-zenith direction (angle B).*

*Figure 6: The blocking effect of the base of the spacecraft on the solar-panel. This example shows partial blocking (the pink area is unseen) but for some observation angles the panel can be completely blocked from view*

Thus, in this model the optical flux that reaches any particular observer from a Starlink may come from one or more of the following spacecraft surfaces and via several scattering types. These contributions (numbered 1 to 4) are shown in Figure 7 and described in the sections below. The angles in Table 1 are relevant to the calculations. The parameters used in the model are listed in Table 2.

| Angles and other factors | Symbol | Description |
|---|---|---|
| Off-base view angle | OBA | The angle between the nadir line at the spacecraft and the direction to the observer |
| Sun depression angle | SDA | The angle that the Sun is below the horizontal at the spacecraft |
| Sun Azimuth | SA | The azimuth of the Sun |
| Panel view angle | PVA | Between the pointing direction of the solar-panel (the normal to the panel) and the direction to the observer |
| Panel Sun angle | PSA | Between the pointing direction of the panel (the mean plane of the panel) and the direction to the Sun |
| Specular beam offset angle | SOA | Angle of the observer from the centre of the specularly reflected beam from the base of the spacecraft |
| Blocking Factor | BF | Fraction of solar-panel front face visible after blocking of observer view by the spacecraft base |
| Range | R | Distance between the spacecraft and the observer |

*Table 1: Angles relevant to the spacecraft position with respect to the observer and the Sun, as calculated from the observation data and epoch*

| Parameter | Symbol | Description |
|---|---|---|
| Pointing mode | PM | Whether the solar-panel is referencing its position to the Sun or the local vertical at the visorsat |
| Panel Offset Angle | POA | Panel offset angle from the vertical at the visorsat |
| Fold Angle | FA | Angle between the folds of the panel and the mean plane of the solar-array |
| Visor Efficiency | VE | Correction for the efficiency of the visor shadowing the base of the spacecraft (as a function of Sun elevation). This includes the fraction of the base affected by the visor |
| Panel reflectivity | a | Parameterised inherent diffuse reflectivity of the bright-face of the Solar-panel (combination of its material properties and the Sun's brightness) |
| Panel rear face emission | b | Parameterised inherent diffuse emission of the dark-face of the Solar-panel |
| Base diffuse reflectivity | c | Parameterised inherent diffuse reflectivity of the base of the spacecraft |
| Base specular reflectivity | d | Parameterised inherent specular reflectivity of the base of the spacecraft |
| Base specular beam dispersion | e | Parameterised angle of dispersion of the specular reflected beam from the base of the spacecraft |
|  | f | Constant used in calculation of apparent magnitude. This is a multiplicative factor that converts a, b, c to visual magnitudes. A value of $7.2^m$ is used. |
| Predicted visual magnitude | $m_{visorsat}$ | The output visual magnitude prediction from the model |

*Table 2: Table of parameters used in the model*

Contribution 1: Diffuse reflection from the front surface of the panel

The front surface of the solar-panel reflects the Sun to the observer if that surface (or any part of it) is facing the observer. Observations of the full illuminated face of the solar- panel on spacecraft in the original 'trains' were around $2.0^m$ (but could be as bright as $-2.0^m$).

At least a significant part of the reflection from the panel is diffuse rather than specular. This diffuse component is here modelled as Lambertian so the power from panel is proportional to the cosine of the Panel View Angle, multiplied by the cosine of the Panel Sun Angle (the Solar flux). Specular reflection of solar flux from the solar-panel may occur. However, in this model the panel is never pointing below the horizontal so that specular beam will never be directed towards (or close to) the Earth.

Contribution 2: Diffuse reflection from the back surface of the panel

The back-surface of the panel, facing away from the Sun, may not be absolutely dark. Observations by the author of Starlinks in their lower parking orbit with the sunlit solar-panel facing away from the observer showed brightness around $6.0^m$. This optical flux, if it exists in the operational configuration, might be due to the illuminated edges of the panel and/or gaps between the panel sections leaking sunlight to the rear side.

In a similar way as Contribution 1, this element is proportional to the negative of the cosine of the Panel View Angle, PVA (PVA is >90° in this case).

Contribution 3: Diffuse reflection from the base of the panel

This is the contribution to brightness that has been assessed in the previous models of Starlink brightness referenced above. The presence of the visor makes it more complex as the amount of exposed antenna surface varies with the elevation of the Sun. Effectiveness is defined here as how much of the bright antenna surface is shadowed. It is assumed here that the visor is designed to be most effective when the Sun is just emerging or appearing behind

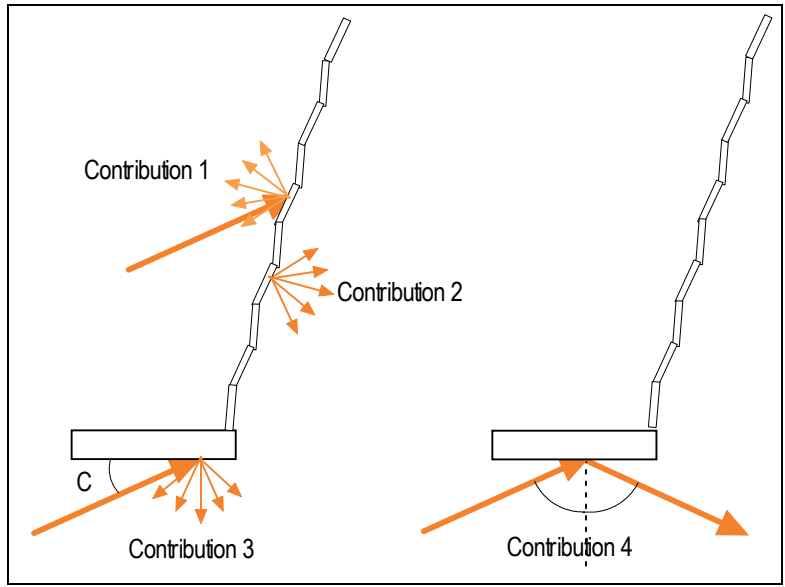

*Figure 7: Possible contributions to the optical flux reflected from a visorsat, as considered in the mathematical model*

Earth, that is when the Sun is 23° below the horizonal at the spacecraft height of 550km. This appears to be a key requirement because if it were not the case, a visorsat in the zenith would get brighter as the night progressed and the Sun descended (until finally the visorsat is in eclipse), the opposite of what is desired to reduce the effect of the Starlinks on astronomical observations.

Reflected sunlight from exposed part of the antennas is modelled as Lambertian diffuse reflection, quantified as the product of the cosine of the off-base view angle and the sine of the Sun depression angle. The effect of the visor was simulated by simple graphics that represented the visorsat base design and the visor as shown in Figure 1. The shadow cast on the base was generated as a function of Sun angle below the base (right panel, 5° to 23°, top to bottom).

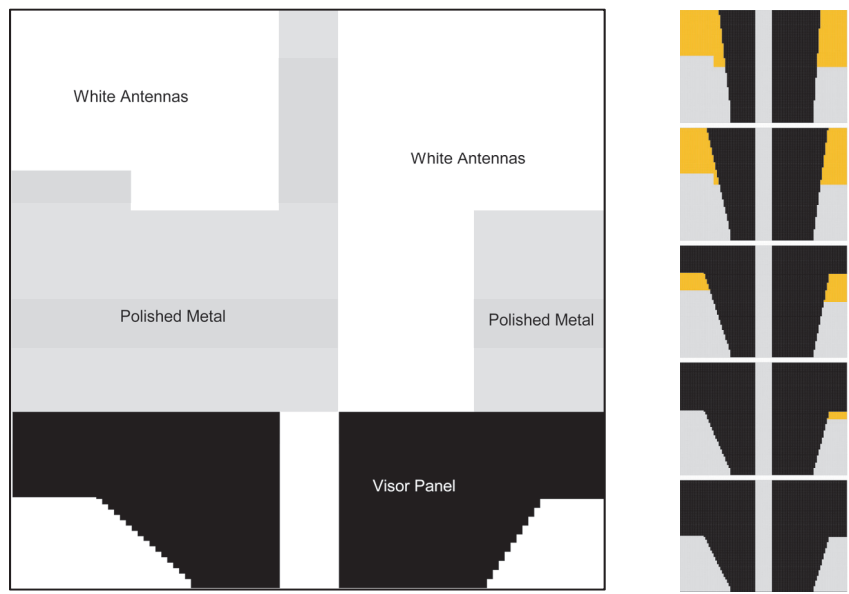

*Figure 8: The visor and base simulation, left. On the right, the shadow cast by the visor as the Sun angle on the panel increases, with illuminated antenna areas shown in orange. The lowest panel shows the antenna panels completely shaded at a Sun angle of 23°*

Convolving the amount of sunlit antenna panel with the amount of sunlight (sine of the Sun angle) gives the curve of scattered sunlight shown in Figure 9 (blue line). There will be no diffuse reflection from those parts of the base in shadow. Some parts of the base are always outside the shadow of the visor as can be seen in the SpaceX diagram (Figure 1) and diffuse reflection for this element will be proportional to the sine of the Sun Angle (red line). The model uses the addition for the two for the flux from the base, adjusted by the Off Base Angle for the observer's view angle. This gives a flux that is flat between Sun angles 10 and 23°, consistent with the observations. Thus, the apparent magnitude of a visorsat in the zenith does not change greatly as the Sun descends.

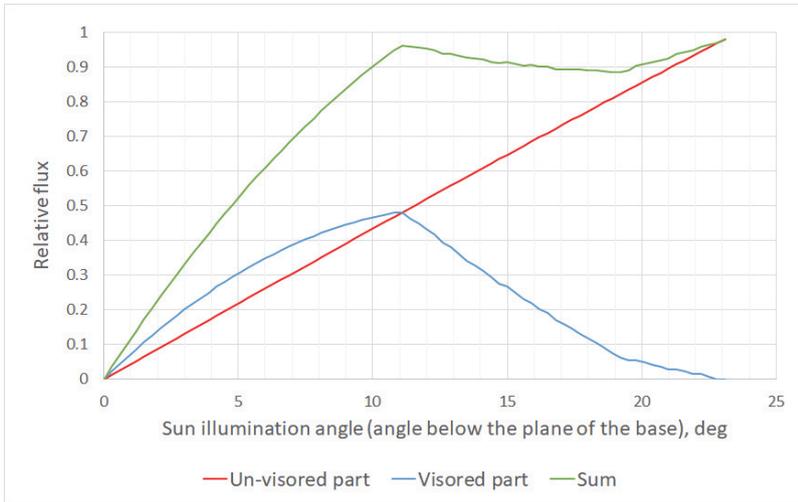

Figure 9: Relative reflected solar flux from fully-illuminated (un-visored) part of base (red), 'visor'-shadowed antenna panels (blue) and total (green). The visor become more efficient at large Sun illumination angles. At smaller angles some sunlight leaks under the visor and is reflected from the antenna panels.

Reflection from the deployable visor itself has not been considered in the model on the basis the visor is declared as being black (as can be seen, peripherally, in an existing image). Also, it faces in the same direction as the solar-panel and will be difficult to differentiate from diffuse reflection from the solar-array.

Contribution 4: Specular reflection from the base of panel

As indicated in Figure 1, large parts of the base of the spacecraft have a mirror-like surface which will specularly reflect at least some of the incident solar flux. The centre line of this beam must miss the Earth but the base surface cannot be a perfect mirror so there will be a substantial angular spread of the beam. Observations of Starlinks at low elevation and close to the Sun azimuth can be within 20° of this specular reflected beam and thus would be affected by sunlight in a widely dispersed beam. The centre of the main specular beam would be very bright, if it could be observed (Figure 7 and Figure 10) so even a small fraction of the light dispersed 20 to 40° from the centre of the beam would be sufficient to make a significant difference to the Starlink's apparent magnitude.

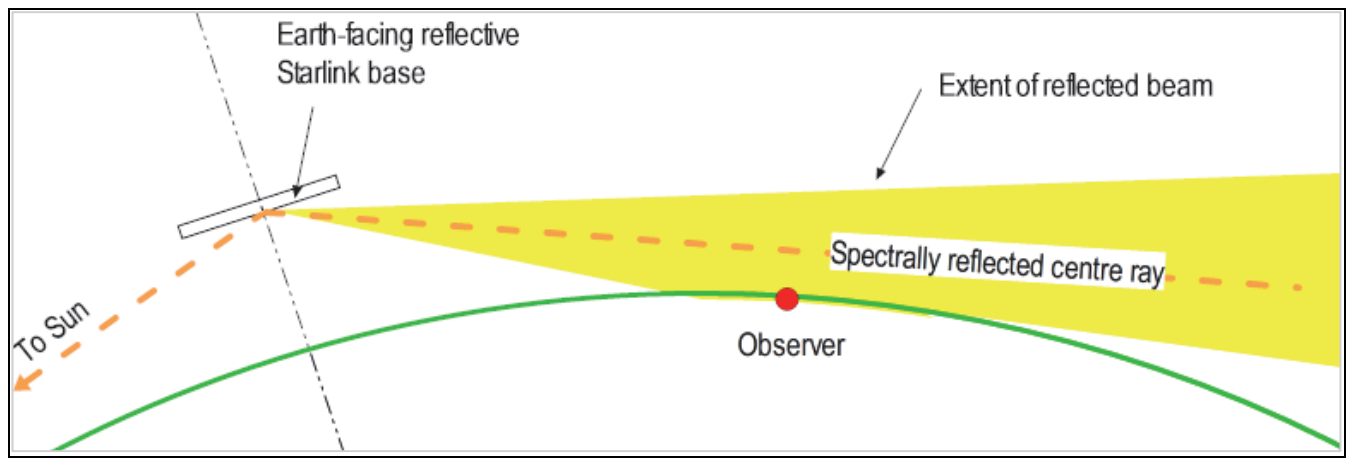

Figure 10: Diagram of the path of the specular beam from a Starlink in the operational orbit at 550km reaching an observer on the ground(not to scale).

In line with methods used in rendering images of modelled objects, the intensity of this beam can be modelled as a large power of the cosine of the angular offset of the observer from the specular beam. Different powers of the cosine in the beam direction and at right-angles seem to be required to explain the observations – the dispersed reflected beam is wider in the plane of the incident and reflected beams.

Figure 11 shows the effect of this observed specular radiation. Observations of the visorsats where the observer view angle is close to the scattered beam show significant magnitude excesses compared to what is expected from the other contributions. This excess can be reasonably modelled by the mechanism described above.

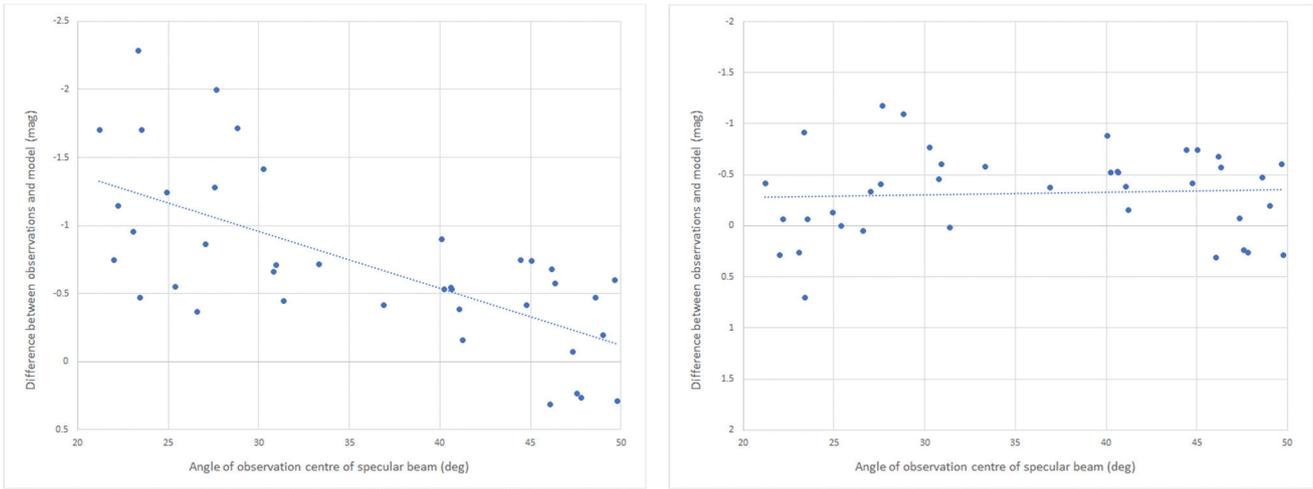

*Figure 11: Observations of specularly reflected sunlight from the base of the spacecraft as a function of angle from the centre of the specular beam. The deviations from the model are before (left) and after application (right) of the correction.*

### 3.3 Summary

On the basis above, the total solar flux (F) from the spacecraft and its final brightness ($m_{visorsat}$) are modelled by equations 1 and 2:

$$F = a \times cos(PVA) \times cos(PSA) \times BF + b \times cos(180 - PVA)$$
$$+ (c \times sin(SDA) \times VE + d \times (cos(SOA))^e) \times cos(OBA) \qquad eq.1$$
$$m_{visorsat} = -2.5 \times \log_{10}(F) + f \qquad eq.2$$

Notes:
1. PVA and PSA take into account PM, POA and FA, as well the geometry of the spacecraft with respect to the Sun and to the observer.
2. Details of the use of the Fold Angle are omitted from the formula for clarity.

## 4 Results from the model

### 4.1 General Behaviour of the model

The model was used to calculate a predicted brightness of the spacecraft for each of the observations, using the observational data and model parameters listed in Table 1 and Table 2. A total of 131 observations made on 66 different visorsat spacecraft were used in this analysis.

The characteristics of the solar-panel brightness measured from ground observations of train spacecraft were used to seed the model and give some additional constraints as discussed below.

The best fits of the model to the data gave an r.m.s. residual of $0.4^m$ that was considered reasonable. The best fit angle of the panel to the spacecraft local vertical was 25° (away from the Sun direction). A plot of the fit residuals when obtaining those parameters is in Figure 12.

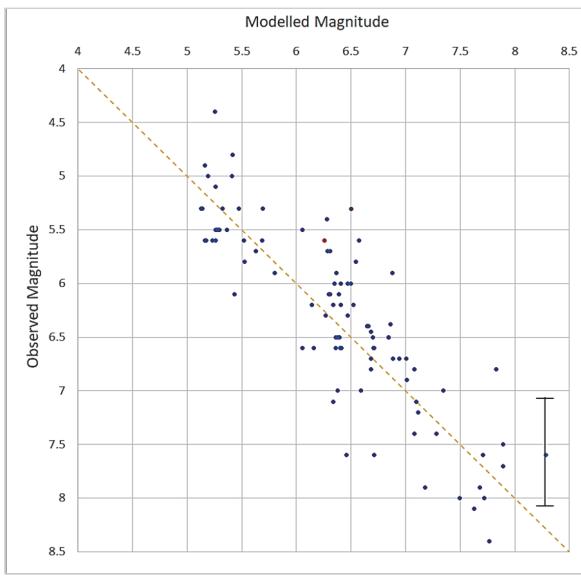

*Figure 12: Fit residuals – the observed visual magnitudes are plotted on the y-axis and the modelled magnitude on the x axis. A typical error bar of each measurement is shown.*

These parameters were used to derive a map of the spacecraft brightness across the sky for any solar elevation angle. An example map is shown in Figure 13 for a Solar depression angle (at the observer) of 15° below the horizon (halfway through astronomical twilight). The positions of the total observation-set on the sky are shown as triangles. The observation azimuths are adjusted for their different Sun azimuths to indicate the sky coverage achieved.

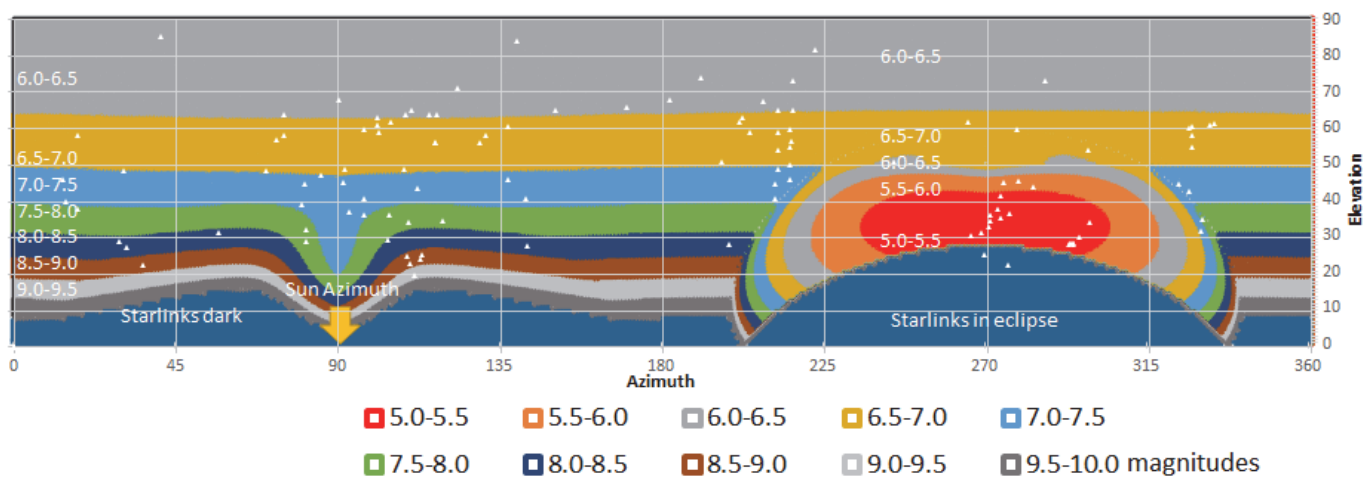

Figure 13: The modelled apparent magnitude of the visorsat across the sky using best-fit parameters. This plot is for the pointing mode with the solar-panel at a fixed angle with respect to the local vertical at the spacecraft. The observation set used in the fit is shown as triangles. Here the modelled solar azimuth is 90° and depression angle 15°.

The following comments are made.

1. The general fit to observations is good, but there are clearly a number of outliers. This may be due to poorer than average observer magnitude estimates or real differences of attitude control methods between different spacecraft. As mentioned above, observations of 66 different spacecraft have been used.

2. There is a complex brightness pattern in the anti-Sun direction, due to visibility of the bright side of the panel and the partial blocking effect of the spacecraft body. The visorsats are considerably brighter in this area of the sky (termed here the 'bright patch') and apparent magnitudes of $5.0^m$ and brighter have been recorded. The area of the sky where the solar-panel is modelled as brighter than the base of the spacecraft is shown in Figure 14. Figure 15 shows the modelled visorsat apparent magnitude for two azimuths.

3. The modelled visual magnitude of the visorsats in the zenith is $6.2^m$ for a Sun depression angle of 15°. This is in reasonable agreement with observation.

4. The contribution of the spacecraft body blocking the view of the solar-panel is vital. If this effect is removed from the model, the size of the bright patch in the anti-Sun increases very significantly. This was mentioned by SpaceX.

5. The effect of the Sun reflecting specularly from the base of the spacecraft in the pro-Sun direction is significant but limited to low elevations near the Sun's azimuth.

6. In the pro-Sun direction the visorsat is modelled to be dark below about 20° elevation, for a Sun depression angle of 15°. This is due to the Sun rising above the plane of the spacecraft base and hence no bright surface is visible.

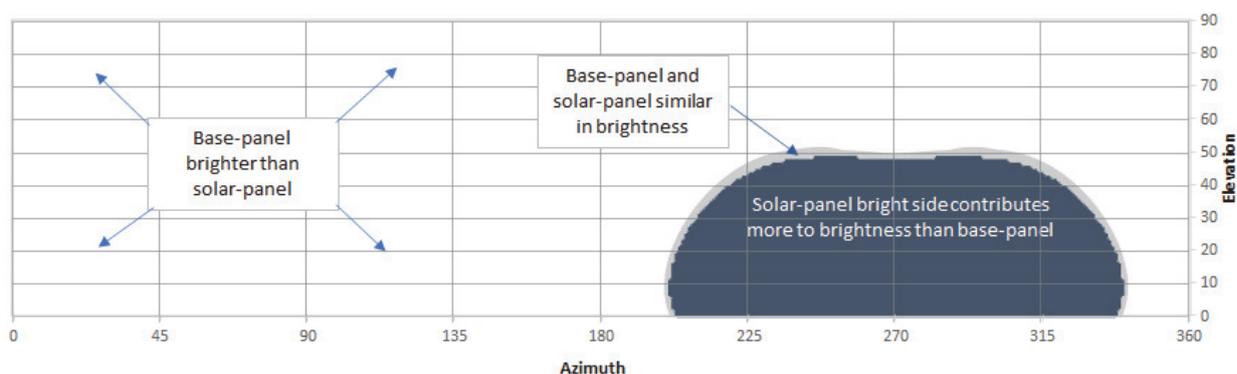

Figure 14: The part of the sky where the solar-panel front face is brighter than the base of the spacecraft, again for a modelled solar azimuth of 90°

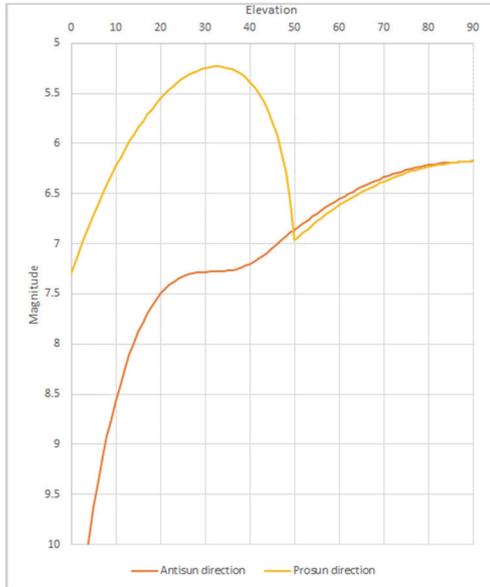

*Figure 15: Modelled apparent magnitude of the visorsat design from horizon to zenith for the pro-Sun and anti-Sun azimuths. The effect of the reflection by the solar-panel refection in the anti-Sun direction is clear. The smaller effect on the pro-Sun curve at elevation 20° is specular reflection from the spacecraft base*

## 4.2 More detailed results of the modelling

The following issues were examined in more detail using the model.

1 - Fold angle of the solar-panel: The observations are not sensitive to the angle of the folds in the solar-panel, so this is set to zero in the model. Many more observations might lead to some constraints on the value of this parameter, but it does not seem to be important to the appearance of visorsats.

2- Brightness of the back-surface of the panel: The original results in Cole (2020a) were that the rear face of the panel was significantly bright at some angles, possibly due to sunlight leaking through gaps. However, the analysis reported here has not confirmed that. The model suggests that the light from the rear of the solar-panel is less than 20% of that from the base (Figure 16).

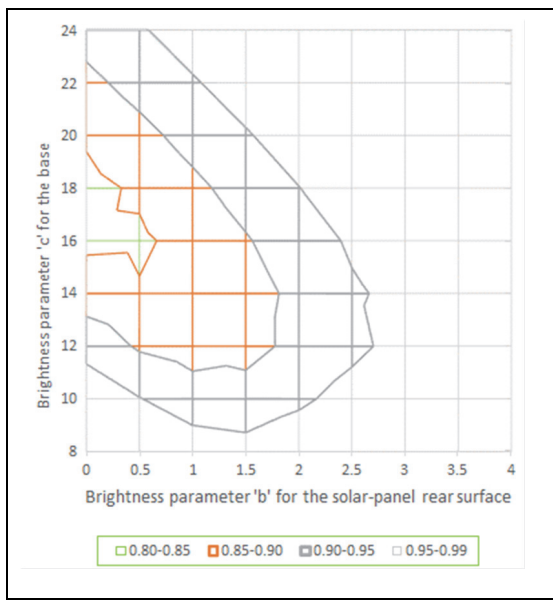

*Figure 16: Confidence contours for the modelled brightness of the rear surface of the solar-panel and the base of the spacecraft. The 95% confidence contour is the grey area. This indicates the optical flux from the rear of the panel is less than 20% of that from the base of the spacecraft.*

3 - Panel pointing referenced to the Sun or local vertical: The fit of the data is better when the panel is modelled as fixed with respect to the local vertical, rather than tracking the Sun with a fixed elevation offset. However, that latter mode of operation is not excluded by the data.

4 - Trade-off between solar-panel surface brightness and tilt angle from the vertical: The data is consistent with a large range of panel offset angles from the vertical, against the inherent brightness of the panel surface (Figure 17). The offset angle can be between 14 and 34°, at the 95% confidence level. The brightness of the panel when seen during Starlink operations at heights lower than 550km, when the solar-panel is pointing directly at the observer, suggest the brightness of the panel is above 90 (in the parameterisation used here) and therefore the panel offset angle is around 26°.

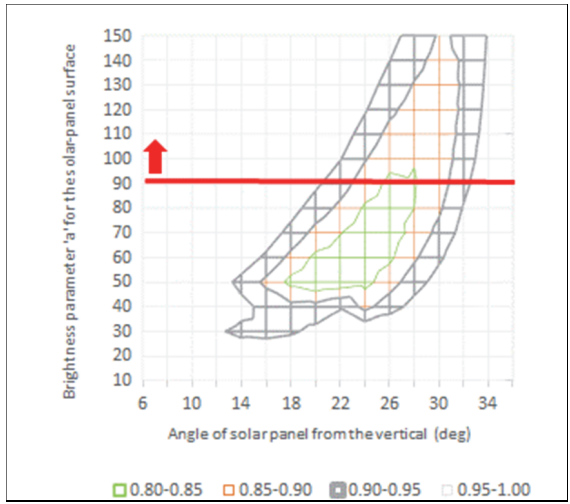

Figure 17: Confidence contours for the solar-panel offset angle against panel surface brightness. The 95% confidence contour is the grey area. The red line indicates a constraint on the brightness from other observations.

5 – Sensitivity of the brightness in the anti-Sun bright patch to panel angle: The brightness of a visorsat in this bright patch is very sensitive to the angle of the panel to the vertical. So, if a particular visorsat uses a different panel angle setting on some or all passes through the 'visible' part of its orbit then it can appear brighter than the nominal model would suggest. There is some suggestion in the observations of some much brighter appearances in this part of the sky. Using the model, the visorsat brightness will change by +/- $0.5^m$ for a panel angle change of +/-6°.

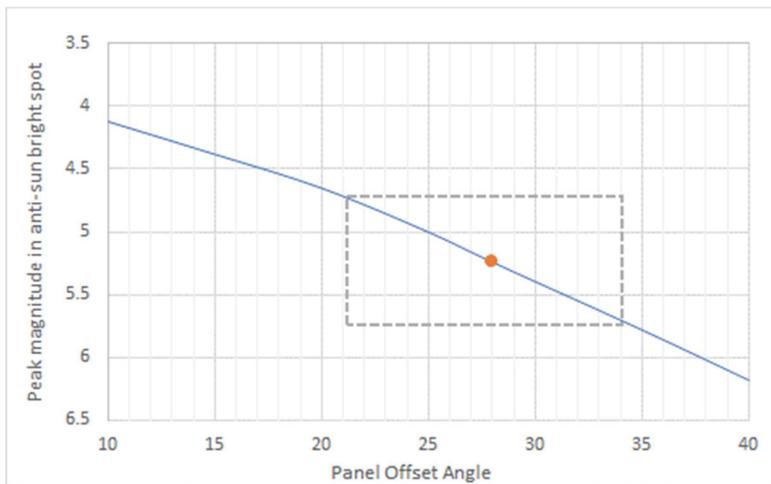

Figure 18: The trend of maximum brightness in the anti-Sun bright patch as a function of angle of the solar-panel to local vertical. The box indicates a +/-$0.5^m$ change in brightness around the nominal panel angle of 28°.

### 4.3 Fitted values of parameters

The best fitted values of the model parameters for the late-2020 dataset are listed in Table 3.

| Parameter | Symbol | Value | Notes |
|---|---|---|---|
| Pointing mode | PM | | Offset from local vertical at the spacecraft |
| Panel Offset Angle (deg) | POA | 24 | Away from Sun azimuth (correlated to a) |
| Fold Angle (deg) | FA | 0 | A non-zero value did not improve the fits |
| Visor Efficiency | VE | | Modelled as per section 3.2 |
| Panel reflectivity factor | a | 70 | Fitted to data (correlated to POA) |
| Panel rear face emission | b | 0 | Non-zero values not required by data |
| Base diffuse reflectivity factor | c | 12 | Fitted to data |
| Base specular reflectivity factor | d | 90 | Fitted to data |
| Base specular beam dispersion (vertical and horizontal) | e | 24 and 52 | Exponents on cosine of angles off-beam centre in vertical and horizontal directions |

Table 3: Best fitted values of the model parameters for the late-2020 dataset.

### 4.4 Change of the appearance of the visorsat at different stages of twilight

A number of plots from the model for a range of solar depression angles are given in appendix 1, showing how the view evolves as twilight progresses.

# 5 Appearance of Pre-visorsat Starlinks

SpaceX launched over 400 operational model Starlinks before the visor feature was developed. The bright antenna panels on the base are thus fully exposed to Sunlight and hence can diffusely reflect this light toward the ground. Earlier spacecraft are brighter than visorsats when at 550km and are relatively easy naked-eye objects in the zenith, though would not be readily noticed by observers not looking for them.

From an image (Figure 15) taken from the ground (Vandebergh 2020b) it appears these spacecraft are rotated around the nadir-zenith axis (presumably only when close to the terminator and visible from the ground) so the panel is edge-on to the Sun and hence much fainter than if it were face-on.

This is an operational method of controlling the overall brightness since there is no method of reducing the brightness of the Earth-facing base of the spacecraft. This rotation manoeuvre is not available to the visorsats as the visor must remain pointing to the Sun-azimuth for it to work. To date no similar images of visorsats in operational orbit have been published.

Since the pre-visorsats have the same polished metal base, they appear to display the same specular-reflection related brightening when low in elevation and close to the Sun azimuth, as discussed for visorsats in section 3.2, contribution 4.

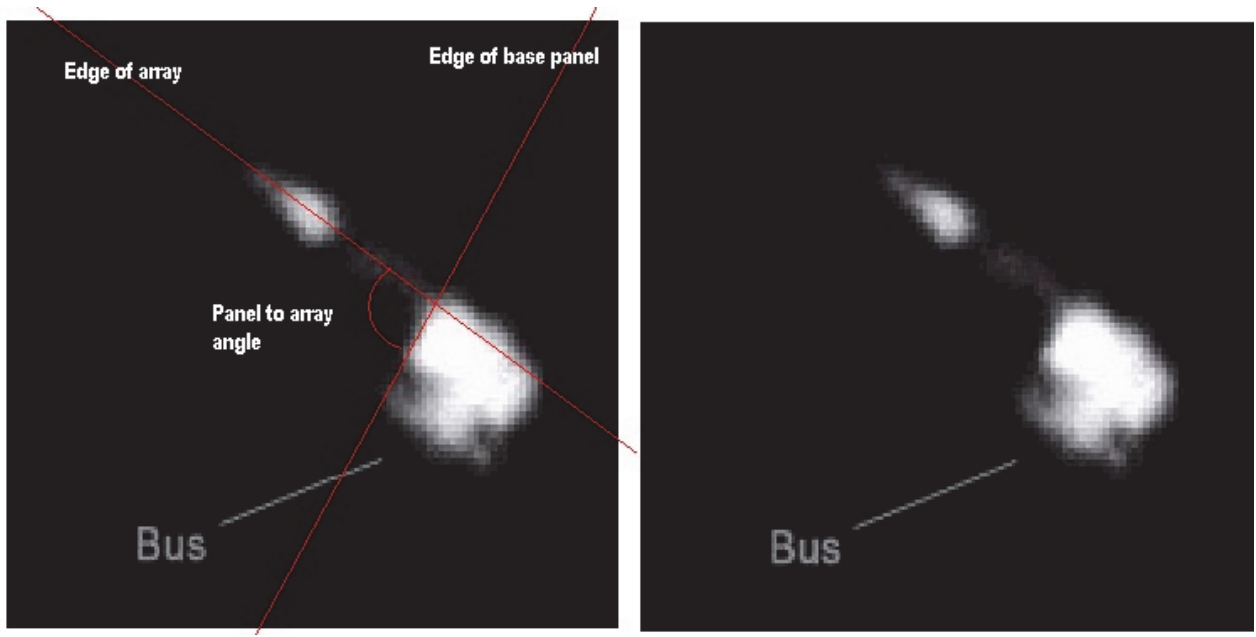

*Figure 19: Ralf Vandebergh's image of pre-visorsat Starlink-1084 at 550km height. The bright solar-panel should be facing towards the observer (if it were Sun-azimuth pointing) but the spacecraft has been rotated to control the brightness.*

# 6 Update – June 2021 observations and model parameters

A further set of observations was made by the Author in June 2021 from a location in SW England at latitude 50.6°. Since late 2020 many more visorsats have reached operational orbit so it is now much easier to acquire multiple observations under the same conditions. The aim of the new observations was to confirm if the bright patch around the anti-Sun direction still existed and if the change of lighting conditions from those of winter to of mid-summer made any difference. The following was immediately apparent from the results:

1. The bright patch in the anti-Sun direction still exists and has become larger and brighter. Visorsats were observed at magnitudes 4-4.5$^m$ and were relatively easy naked-eye objects.

2. The visorsat brightness at zenith and towards the Sun was unchanged, though not deeply investigated.

The model was adapted to fit this data by adjusting the angle of tilt of the solar panel to -5° to the vertical, that is inclined to the Sun azimuth, rather than >15° away from the Sun which was the indication from the earlier dataset (Figure 17). For observations close to the anti-Sun azimuth, the two datasets are shown in Figure 20, with the best fits for the late 2020 and June 2021 sets. The extension of the bright patch almost to the zenith is apparent in the June 2021 set.

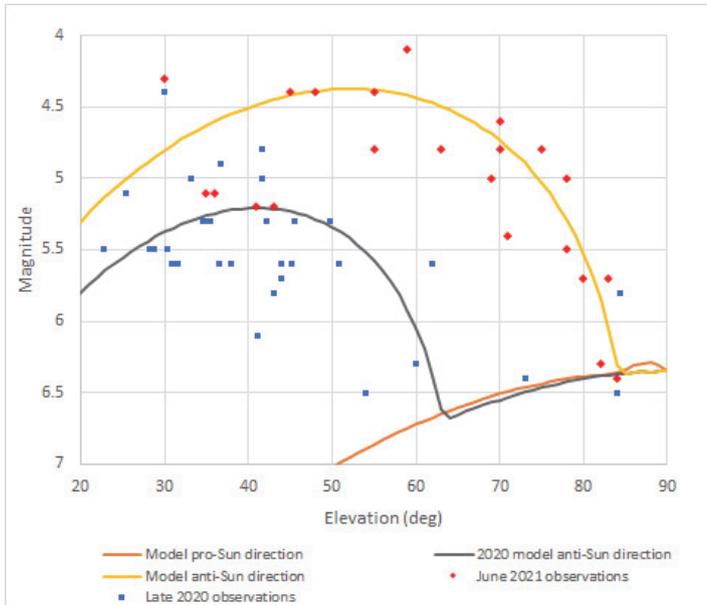

Figure 20: The late 2020 and June 2021 datasets of visorsat magnitudes near the anti-Sun azimuth with the modelled curves using different solar panel angles with respect to local vertical at the spacecraft, as discussed in the text

As indicated by SpaceX (SpaceX 2020) the angle of the panel is controlled to reduce the brightness of the visorsat from the ground. This is only needed when the spacecraft is potentially visible, that is when the Sun is below the horizon on the ground under the spacecraft and the spacecraft is not in shadow. Any off-pointing of the array from the Sun reduces available power and this may be critical for some orbits, for example where the spacecraft spends an extended part of the orbit where that ground visibility criteria is met. Therefore, there may be operational reasons for the previous rule to be amended.

Astronomically, the area of the sky close to the Sun during twilight has been stated as critical for Near-Earth Object studies, and this is unaffected by the changed behaviour discussed here. Also, mid-summer at high latitudes is (arguably) not an important astronomical season since the skies are bright all through the night. There is also evidence that SpaceX respond to complaints and if there are none from certain areas at certain times of the year then a simpler operational rule may be applied.

On this basis, a corresponding map of Starlink brightness over the sky is shown in Figure 21, to be compared with Figure 13. This is initial work, to be confirmed with further observations.

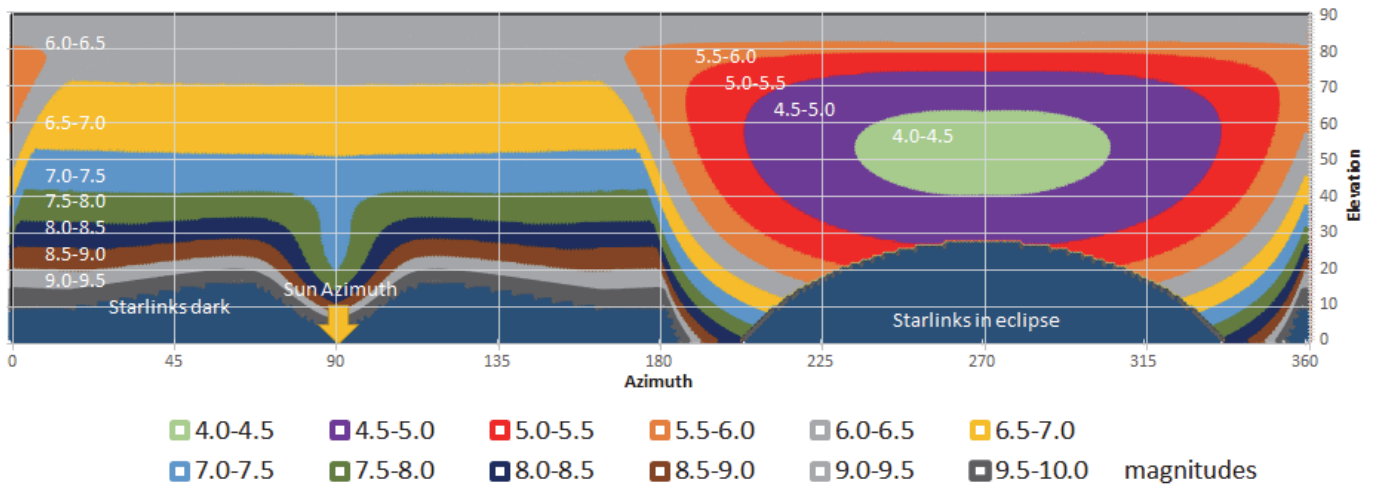

Figure 21: The modelled apparent magnitude of the visorsat across the sky using updated best-fit parameters for the June 2021 dataset. This plot is for the pointing mode with the solar-panel at a fixed angle with respect to the local vertical at the spacecraft, in this case 5° towards the Sun-azimuth. The modelled solar azimuth is 90° and depression angle 15°.

## 7 Conclusions

The main conclusions of this work on the model are as follows:

1. The brightness of a visorsat in the zenith is 6.2$^m$ for Sun elevation of -15° and changes little for Sun elevations from -9° to -23°. This stability is due to the way the visor works as the Sun elevation changes.

2. There is a complex brightness pattern in the anti-Sun direction due to visibility of the bright side of the panel and the blocking effect of the spacecraft body. Apparent magnitudes as great as $4.0^m$ have been observed over a limited sky area in recent observations. There is no indication that either the pre-visorsat starlinks or Darksat display this bright behaviour, presumably for the reasons stated in the previous section.

3. There is evidence that the visorsat bright behaviour in the anti-Sun direction is not consistent with time. From initial observations the brightness behaviour at zenith and towards the Sun azimuth appears stable.

4. The visorsats are marginally brighter, low in the sky on similar azimuths to the Sun (so at high phase angles). This also seems to be the case for pre-visorsat Starlinks so is not due to the visor.

5. On this basis the visorsat design is only marginally fainter at zenith than the Darksat design, which has been observed at high elevation at $5.8^m$ by the author (Cole 2020) and by others. The pre-visorsat Starlinks are $4.6^m$ under similar conditions.

This report is being updated as work proceeds. Work continues to make and obtain more observations to allow further detailing of the model.

-13-

# REFERENCES

Beskin, G.M. et al. Wide-field optical monitoring with Mini-MegaTORTORA (MMT-9) multichannel high temporal resolution telescope. Astrophys. Bull. 72, 81–92 (2017). https://doi.org/10.1134/S1990341317030105

Cole, R. E. "A Sky Brightness Model for the Starlink 'Visorsat' Spacecraft - I" (2020a), Research Notes of the American Astronomical Society, 4, 10, https://iopscience.iop.org/article/10.3847/2515-5172/abc0e9

Cole, R. E. "Measurement of the Brightness of the Starlink spacecraft named "DARKSAT" (2020b), Research Notes of the American Astronomical Society, 4, 3 https://iopscience.iop.org/article/10.3847/2515-5172/ab8234

Dark and Quiet Skies for Science and Society, "Impact of satellite constellations workshop 5-9 October" (2020) https://owncloud.iac.es/index.php/s/WcdR7Z8GeqfRWxG#pdfviewer

Hainaut, O. R., and Williams, A. P. "Impact of Satellite Constellations on Astronomical Observations with ESO Telescopes in the Visible and Infrared Domains." Astronomy & Astrophysics (2020): arxiv.org/abs/2003.01992

Mallama, A, private communication (2020a)

Mallama, A, Starlink L10 VisorSat Magnitudes (2020b) http://www.satobs.org/seesat/Oct-2020/0073.html

Mallama, A, "Starlink Satellite Brightness Before VisorSat" (2000c) https://arxiv.org/ftp/arxiv/papers/2006/2006.08422.pdf

Respler, J, "Starlink Obs 20-35 & 20-55" (2020) http://www.satobs.org/seesat/Oct-2020/0003.html

SpaceX, "Starlink Discussion National Academy of Sciences April 28, 2020" (2020) https://www.spacex.com/updates/starlink-update-04-28-2020/

Vandebergh, R, tweet (2020a) https://twitter.com/ralfvandebergh/status/1250772537375670278?s=20

Vandebergh, R, private communication (2020b

# Appendix 1:

A series of predicted sky maps from the model, showing how the modelled visorsat magnitudes evolve as the Sun descends in the sky.

1. Sun depression angle 9° (middle of nautical twilight)

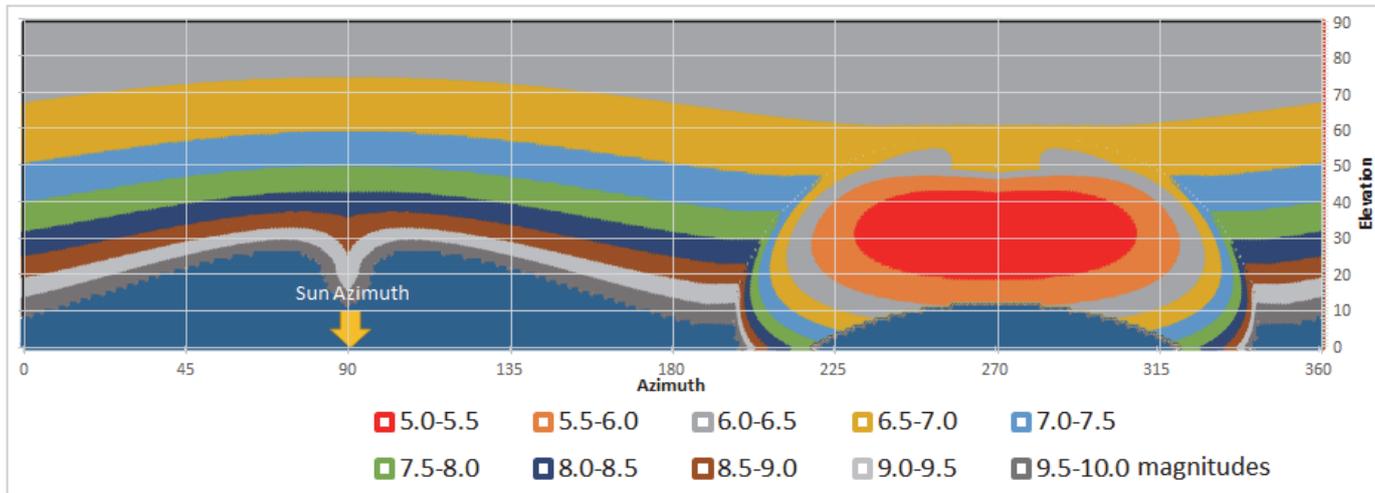

Notes:
   a. Visorsat visual magnitude at zenith is 6.3$^m$
   b. The blue patch at azimuth 270° is the shadow of the Earth at 550km (satellites in eclipse)

2. Sun depression angle 12° (start of astronomical twilight)

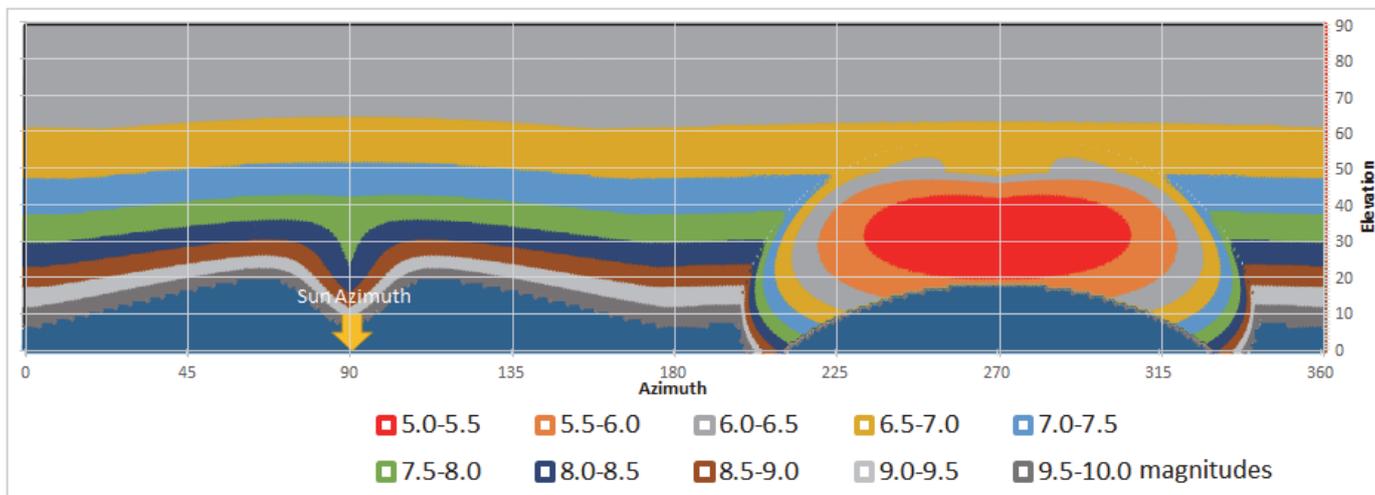

Notes:
   a. Visorsat visual magnitude at zenith is 6.1$^m$

3. Sun depression angle 18° (end of astronomical twilight)

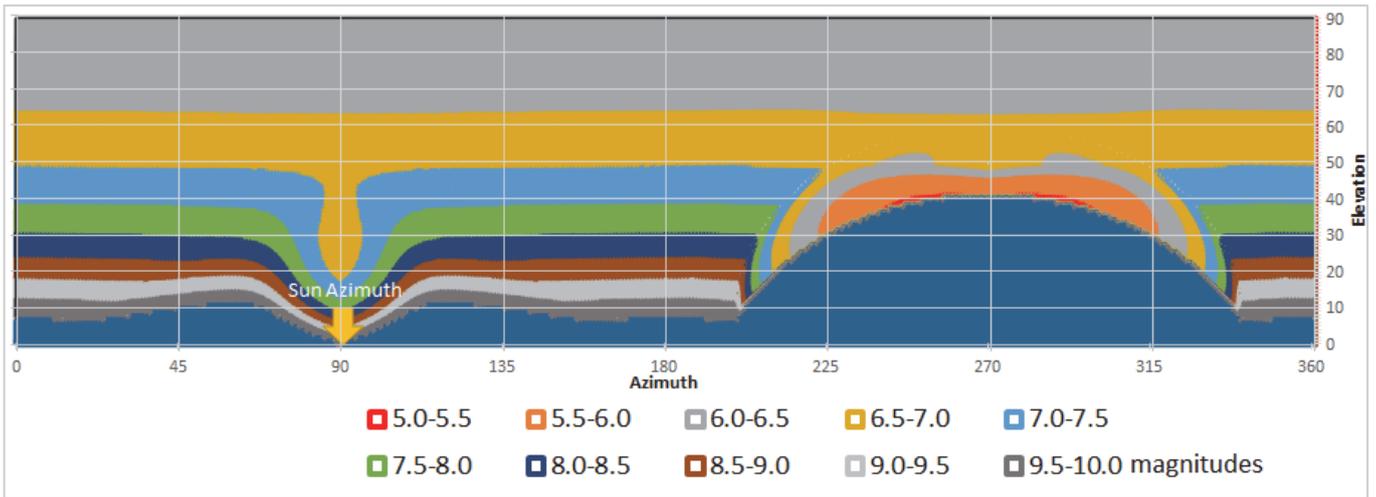

Notes:
a. Visorsat visual magnitude at zenith is 6.2$^m$

4. Sun depression angle 23°

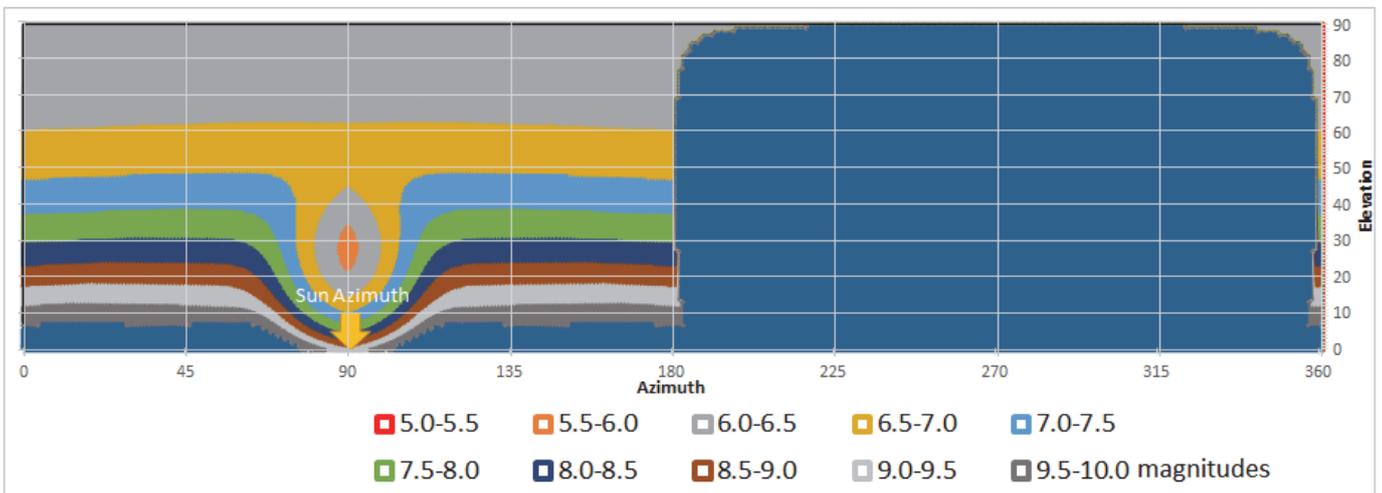

Notes:
a. Starlinks at zenith will be entering or leaving eclipse.
b. Visorsat visual magnitude at zenith is 6.1$^m$

5. Sun depression angle 30°

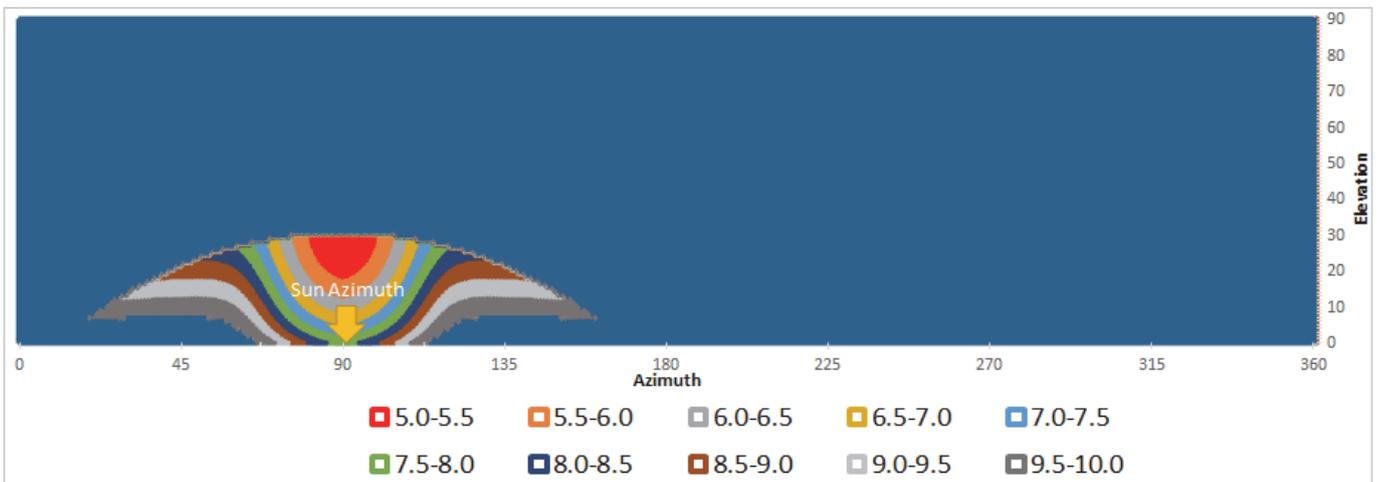

6. Sun depression angle 36°

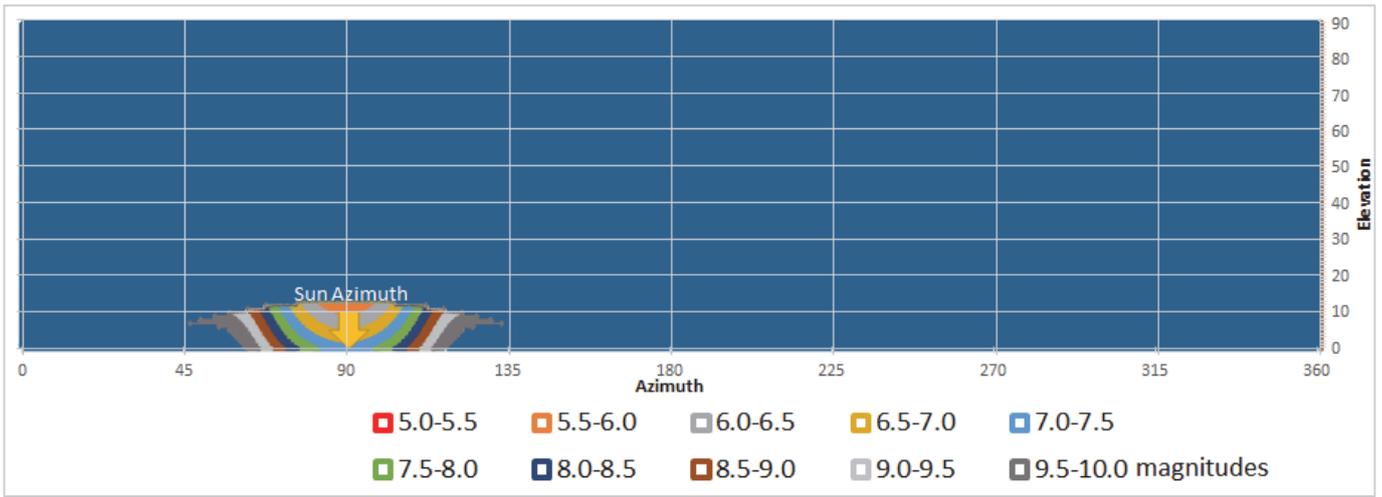

Change record

| Version | Date | Section changed | Changes |
|---|---|---|---|
| 1 | 31/10/20 | | First version |
| 2 | 1/11/20 | 4.2 | Added issue #5 to the list of detailed issues discussed. |
| 3 | 10/7/21 | 4.3 | Added section with fitted parameter values |
| | | 5 | Figure 19 updated to better version |
| | | 6 | Added section on June 2021 observations |
| | | 7 | Updated conclusions |